\documentstyle{article}
\newcommand{\beq}{\begin{equation}}
\newcommand{\eeq}{\end{equation}}

\newcommand{\ie}{{\em i.e.\/}}

\begin{document}
\begin{center}
{\Large\bf Inhomogeneous scalar field solutions and inflation}\\[.5cm]
S. E. Perez Bergliaffa \footnote{E-mail: santiago@lafex.cbpf.br} 
and K. E. Hibberd\\
Centro Brasileiro de Pesquisas Fisicas\\
Rua Xavier Sigaud, 150, CEP 22290-180, Rio de Janeiro - RJ\\
Brazil
\end{center}
\begin{abstract}
We present new exact cosmological inhomogeneous solutions for gravity coupled to a scalar field 
in a general framework specified by the parameter $\lambda$. The equations of motion (and 
consequently the solutions) in this framework 
correspond to either low-energy string theory or Weyl integrable spacetime according to the sign of $\lambda$.
We show that different inflationary behaviours are possible, as suggested by the study
of the violation of the strong energy condition.  Finally, by the analysis of certain curvature scalars
we found that some of the solutions may be nonsingular.

\end{abstract}


\section{Introduction}

The inflationary paradigm was first introduced by Guth \cite{guth} and
solves, without the necessity of
imposing highly special conditions on the initial state of the universe, 
some of the problems of the
standard cosmological model. However, the homogeneity 
and isotropy problem still deserves some attention. This is due to the fact that 
a general proof of
the naturalness of inflation is still to be derived. That is to say, it has not been shown that 
inflation will take place in any spacetime, independently of its symmetries.
Since Gibbons and Hawking \cite{gib1} and Gibbons and Moss \cite{gib2}
have stated the ``cosmic no hair'' conjecture, 
some progress has been  made by Wald
\cite{wald} in the case of Bianchi models.  There has also been contributions from Jensen and Stein-Schabes
\cite{stein}, and Stein-Schabes and Barrow \cite{barrow} for a class of inhomogeneous models
which was
quite close to spatial homogeneity. In all of these cases, 
inflation was driven by a positive cosmological constant as the conjecture 
requires. 

In the more general case of a scalar field powering the expansion, 
Collins and Hawking \cite{hawking} studied whether homogeneous models can approach 
isotropy. They found that only an insignificantly small number of these models 
can isotropize. Morover, there are some 
``no go'' results for a large class of scalar field potentials \cite{heusler}.
Byland and Scialom \cite{byland} studied the issue of isotropization and inflation in 
Bianchi I, Bianchi III and Kantowski-Sachs models. They found that isotropy can be reached
without inflation for Bianchi I and that if inflation takes place then isotropy
is always reached, confirming previous results by Burd an Barrow \cite{burd}.

In spite of these advances, whether or not inflation is generic is
still an open issue either in the case of anisotropic or inhomogeneous scalar 
field universes. In particular for the inhomogeneous case some numerical 
and qualitative work has been done on the occurrence of inflation
when initial inhomogeneities are present \cite{iguchi,kurkisuonio,goldwirth}.
However,
only a few exact solutions can be found
in the literature. In a number of papers,
Feinstein {\em et al} \cite{aguirre,fil,fein2} showed that the behaviour of the inhomogeneous
solutions can be diverse. In some cases their solutions inflate and homogeneize, but they do
not isotropize \cite{aguirre}. In \cite{fil} it was shown that only some regions of spacetime 
undergo inflation, while for the geometries discussed in \cite{fein2} inflation never occurs.

The scalar field which is a basic ingredient for inflation,
arises naturally at in least three
different contexts: Kaluza-Klein theories \cite{bailin}, low-energy string theory \cite{witten} and 
Weyl's theory in its integrable version (WIST) \cite{novello}. 
As we shall see later, a salient feature of WIST is that for a certain range of values of a 
parameter $\lambda$, the equations of motion coincide with those obtained in the low-energy limit of 
string theory in the so-called Einstein frame, 
neglecting all matter fields but the dilaton 
\footnote{In turn, these are the equations for 
General Relativity plus a minimally coupled scalar field.}
\cite{barrowkunze}.
Consequently, our aim is to derive and analyze the properties of inhomogeneous 
cosmological solutions of the equations of motion in the general setting
of WIST. 
In so doing, we will obtain new solutions for low energy string theory in
the case mentioned above and 
solutions with no counterpart in General Relativity (GR), \ie$~$
unique to WIST.  

In this work we shall use a model which generalizes Bianchi I 
cosmologies by the addition of some inhomogeneity in one dimension. 
These models are describable by the generalized Eistein-Rosen metric \cite{einsteinrosen}.
The origin of the inhomogeneity may be related 
to the presence of a background of primordial gravitational waves, which 
may have played a role in the early stages of the Universe \cite{aguirre}. 
The scalar field of the model is under the influence of a Liouville 
type potential, which arises 
as an effective potential in many unification theories \cite{halliwell}. 
Also this potential might give rise to a power law-inflation.  It
has been suggested by M\"uller {\em et al} \cite{muller} that this 
type of inflation is an attractor in the space of solutions of Einstein's 
equations in the limit $t\rightarrow\infty$ for the case of inhomogeneous spacetimes 
with a minimally coupled scalar field.

The structure of the paper is as follows. In Section 2 we give a short review of the basics 
of WIST integrable geometry and in Section 3 we derive the field equations. Sections 4
and 5 are devoted to two different types of solutions for the generalized Einstein-Rosen geometry. 
In Section 6 we discuss 
whether the solutions display inflation as indicated by the violation of the strong energy condition (SEC).
We also study their singularity pattern by 
considering certain curvature scalars.
Our conclusions and final remarks are given in Section 7.

\section{Weyl integrable space-time}

Weylian geometry is a generalization of Riemannian geometry in which the
metric and a scalar field are the fundamental objects \footnote{A review of Weylian
geometry can be found in \cite{salimsautu}.}. The difference 
between the two geometries stems from the law of parallel transport, which in the case of 
Weylian geometry is given by
\beq
g_{\mu\nu ;\delta} = \omega_\delta g_{\mu\nu}.
\eeq
The field $\omega_\delta$ governs the variation of the standards of measure according 
to
\beq
dl = \frac l 2\omega_\alpha dx^\alpha.
\eeq
If we impose that the vector field $\omega_\mu$ be a gradient of a certain field $\omega$, that is,
\beq
\omega_\mu = \partial_\mu\omega ,
\eeq
then length variations along a closed path are integrable. The resulting geometrical structure
is called Weyl Integrable Space-Time (WIST) \cite{novello}.
We see therefore that in WIST the scalar field has a purely geometrical 
origin. 

The simplest action we can write for this geometry in a
vacuum is
\beq
S_W = \int (R+\xi \omega^{;\mu}_{\; ;\mu}) \;\sqrt{-g}\;d^4x,
\eeq
where $R$ is the the scalar curvature for the Weylian geometry and $\xi$ is an 
arbitrary coupling constant. The variation of $S_W$ gives the following equations
of motion
\beq
G_{\mu\nu} + \omega_{;\mu ;\nu} - (2\xi - 1) \omega_{;\mu}\omega_{;\nu} + \xi
g_{\mu\nu} \omega_{\; ;\alpha}\omega^{\; ;\alpha} = 0,
\eeq
\beq
\omega^{;\alpha}_{\; ;\alpha}+2\omega_{;\alpha}\;\omega^{;\alpha} = 0.
\eeq
These equations are written in terms of the tensors associated to the Weylian 
geometry. However they can be recast using the corresponding Riemannian tensors 
in the following way \cite{salimsautu}:
\beq
\widehat G_{\mu\nu} = -\lambda (\omega_{||\mu}\omega_{||\nu} - \frac{1}{2}g_{\mu\nu}
\omega^{||\alpha}\omega_{||\alpha}),
\label{field1}
\eeq
\beq
\widehat{\raisebox{-.5ex}{$\Box$}}\omega = 0,
\label{field}
\eeq
where $\lambda = 3-4\xi$. According to the conventions we 
are using for the Riemann tensor, the source-like term coincides with the stress-energy
tensor of a scalar field in a Riemannian spacetime only if the constant $\lambda$ is negative. 
As was stated above, the solutions will depend 
on the parameter $\lambda$, which determines the type of theory 
we are dealing
with. For a negative $\lambda$ we recover GR, while $\lambda>0$ provides us with a
different sector of WIST.
Some solutions of interest in the case of $\lambda <0$
have been mentioned in the introduction. 
Several exact solutions with $\lambda>0$ have been found in WIST. Among them, 
we can cite nonsingular cosmologies \cite{novello},
static and spherically symmetric spacetimes \cite{sasa2},
inflationary solutions \cite{fasasa}, and dilaton electromagnetic Bianchi-I 
cosmologies \cite{sasama}. The coupling of different types of matter to the geometry of
WIST was studied in \cite{sasa}.  In the next section we will derive the expression for the field equations 
(\ref{field1}) and (\ref{field}) in the generalized Einstein-Rosen metric.

\section{Field equations}

In what follows, we shall work with one-dimensional inhomogeneities which are describable by the 
generalized Einstein-Rosen metric. 
These spacetimes admit an Abelian group of
isometries $G_2$ and include Bianchi models 
of types I-VII \cite{einsteinrosen}.
Assuming that the two Killing vectors are hypersurface orthogonal,
the metric of the model is given by the following diagonal form:
\beq
ds^2\; = \; e^{f(t,z)}(-dt^2+dz^2) + G(t,z)[e^{h(t,z)}dx^2 + e^{-h(t,z)} dy^2].
\label{metric}
\eeq
The potential is taken as 
\beq
V(\omega ) = \Lambda e^{-\lambda k \omega(t,z)},
\eeq
where $\Lambda$ and $k$ are arbitrary constants. 
The field equations for the geometry are 
\beq
\frac{\ddot G}{G} - \frac{G''}{G} = 2 e^f V,
\label{feq1}
\eeq
\beq
\frac{\ddot G}{G} + \frac{G''}{G} - \frac{1}{2} \left(\frac{\dot G}{G}\right)^2 - \frac{1}{2}
\left(\frac{G'}{G}\right)^2 - \frac{\dot f\dot G}{G} - \frac{f' G'}{G} + \frac{1}{2}
\dot h^2 + \frac{1}{2} h'^2 = \lambda (\dot\omega^2 +  \omega '^2),
\label{feq2}
\eeq
\beq
\frac{\dot G'}{G} - \frac{\dot G G'}{2G^2} + \frac{1}{2}\dot h h' - \frac{f' \dot G}{2G} 
- \frac{\dot f G'}{2G} = \lambda \dot\omega \omega ',
\label{feq3}
\eeq
\beq
\ddot h - h'' +\frac{\dot G\dot h}{G}- \frac{G' h'}{G} = 0,
\label{feq4}
\eeq
\beq
\ddot \omega - \omega '' +\frac{\dot G\dot\omega}{G} - \frac{G' \omega '}{G} -\frac{e^f}
{\lambda}\frac{\partial V}{\partial\omega}=0.
\label{feq5}
\eeq
We shall find particular solutions to these equations for two different
cases: $G=G(t)$ (which corresponds to the case in which the element of the 
transitivity surface is homogeneous) and $G=G(t,z)$. To solve the equations we shall adapt 
the method presented in \cite{aguirre} for the first case and
the steps given in \cite{fil} for the second case.

\section{G=G(t)}

To solve Eqs.(\ref{feq1})-(\ref{feq5}), we write the scalar field as follows \cite{aguirre},
\beq
\omega (t,z) = -\frac{k}{2}\ln G(t) + \Omega (t,z) .
\label{omega}
\eeq
After substituting Eq.(\ref{omega}) into Eq.(\ref{feq5}) we obtain an expression for $\Omega (t,z)$,
\beq
\ddot\Omega - \Omega '' + \frac{\dot G}{G}\dot \Omega =0.
\label{Omega1}
\eeq
Suppose that the functions $h(t,z)$ and $\Omega (t,z)$ can be split as
\beq
h(t,z) = \Pi (t) + P(z),\;\;\;\;\;\;\;\;\;\;\;\;\;\;\;\;\;\;\; \Omega (t,z) = \chi (t) +\psi (z).
\label{sep}
\eeq
Using  Eq.(\ref{sep}) in Eqs.(\ref{feq4}) and (\ref{Omega1}) we get
\beq
h (t,z) = \Pi (t) +\frac{B}{2}z^2 + Cz, \;\;\;\;\;\;\;\;\;\;\;\;\;\;\;\;\;\;\
\Phi (t,z) = \chi (t) + \frac{E}{2}z^2+Fz,
\label{hphi}
\eeq
where $B,C,E$ and $F$ are arbitrary constants of integration.  Substituting this equation and  the expression for 
$f(t,z)'$ obtained
by differentiating Eq.(\ref{feq1}) into
Eq.(\ref{feq3}) we obtain
\beq
\frac{1}{2}\dot\Pi (Bz+C) -\lambda \dot\chi (Ez+F) = 0,
\label{rel}
\eeq
where $\lambda$ is the constant that appears in the equations of motion.
From Eqs.(\ref{feq1}), (\ref{feq5}) and (\ref{hphi}) we arrive at
\beq
\frac{\ddot G}{G} - \frac{\stackrel{...}{G}}{\ddot G}\frac{\dot G}{G} - K \left( \frac
{\dot G}{G}\right) ^2 + \frac{1}{2}\dot\Pi ^2 - \lambda \dot\chi ^2
+\frac{1}{2}(Bz+C)^2 - \lambda (Ez+F)^2 = 0,
\label{g1}
\eeq
with $K = -\frac{1}{4}(\lambda k^2+2)$. It follows from this equation and Eq.(\ref{rel})
that
\beq 
\dot \Pi = \frac{2\lambda F}{C}\dot\chi .
\label{rel2}
\eeq
Eq.(\ref{feq4}) now gives
\beq
\dot\Pi = \frac{A}{G},
\label{rel3}
\eeq
where $A$ is a constant. 

Finally, after substituting Eqs.(\ref{rel2}) and (\ref{rel3}) 
into Eq.(\ref{g1}) we obtain an equation that governs the evolution
of the function $G(t)$:
\beq
\ddot G^2 G - \stackrel{...}{ G} \dot G G - K \ddot G \dot G^2 + M^2\ddot G + N^2 G^2 \ddot G = 0,
\label{maineq}
\eeq
where the constants $M$ are $N$ are given by
\beq
M^2 = \frac{A^2}{2}\left( 1 - \frac{C^2}{2\lambda F^2}\right), \;\;\;\;\;\;\;\;\;\;\;\;\;\;\;\;
N^2 = \left( \frac{C^2}{2} - \lambda F^2\right).
\eeq
Also the functions appearing in the metric can be rewritten as the following
\begin{eqnarray}
h(t,z) & =  & A\int\frac{dt}{G} + Cz, \\
\omega (t,z)& =  & -\frac{k}{2}\ln G + \frac{CA}{2F\lambda}\int\frac{dt}{G} + Fz, \\
f(t,z) & = & \lambda k\omega + \ln \frac{\ddot G}{G} - \ln 2\Lambda.
\end{eqnarray}
Therefore any solution of Eq.(\ref{maineq}) will determine 
the functions in the metric (\ref{metric}). 

Next we present particular cases of this geometry. Each case is associated to 
a different choice of the function $G(t)$.  A summary of the solutions and constraints on the 
parameters is displayed in Table 1
for $\lambda >0$.

\begin{table}[p]
\centering
\begin{tabular}{||l|l|l||} \hline \hline
$G(t)$& Constraints & Functions \\ \hline \hline
 $e^{\beta t}$ & $\beta ^2 = 2\;\frac{2\lambda F^2 - C^2}{\lambda k^2 + 2},$ & $h(t,z)  =  Cz$, \\ 
 & $\lambda > \frac{C^2}{2F^2},$ & $\omega (t,z)  =  - \frac{k}{2}\beta t +Fz$,  \\
 & $A=0~~$,  $\Lambda > 0$, & $f(t,z)  =  -\frac{\lambda k^2}{2}\beta t + \lambda k F z + \ln\left(\frac
{\beta^2 }{2\Lambda}\right).$  \\ \hline 
 $t^\beta$ & $\beta = -2\;\; \frac{1\pm \sqrt{1+2(\lambda k^2+2)(2\lambda F^2-C^2)}}{\lambda k^2+2},$
       & $ h(t,z)  =  Cz    $, \\
& $\lambda > \frac{C^2}{2F^2}$, & $ \omega (t,z)  =   -\frac{k}{2}\beta \ln t +Fz  $,\\

& $A=0,$ & $f(t,z)  =  -\left(\frac{\lambda k^2}{2} \beta + 2\right) \ln t + \lambda k F z
 + \ln \left[ \frac{\beta (\beta - 1)}
{2\Lambda}\right].$\\ \hline 
  $ \alpha\sinh \beta t$ &   $\beta ^2 = 2\;\frac{2\lambda F^2 - C^2}{\lambda k^2 + 2},$ &
$h(t,z)  =  \frac{A}{\alpha \beta} \ln\left(\tanh\frac{\beta t}{2}\right) + Cz,$ \\
    &$\alpha ^2 =\frac{A^2}{2\lambda F^2 }\;\frac{\lambda k^2 +2}{ 2-\lambda k^2}\;,$ &
 $\omega (t,z)  =  -\frac{k}{2}\ln \alpha + \ln \left[ \frac{(\tanh \frac{\beta t}{2})^\delta }
{(\sinh \beta t)^{\frac k 2}}\right] +Fz,$ \\
    &$\lambda\in \left( \frac{C^2}{2F^2},\frac{2}{k^2}\right),~~    
    \delta = \frac{AC}{2\lambda F\alpha\beta}, ~~ \Lambda>0, $&
   $f(t,z)  =  \ln \left(\frac{\beta ^2\alpha^{-\lambda k^2/2}}{2\Lambda}\right) +
\ln\left[\frac{(\tanh \frac{\beta t}{2})^{\delta\lambda k}}{(\sinh \beta t)^{\lambda k^2/2}}
\right] + \lambda k F z .$\\ \hline 
$\alpha \sin \beta t$ & $\beta ^2 = -2\;\frac{2\lambda F^2 - C^2}{\lambda k^2 + 2},$ &
 $h(t,z)  =  \frac{A}{\alpha \beta} \ln\left(\tan\frac{\beta t}{2}\right) + Cz,$ \\
 &$\alpha ^2 =\frac{A^2(\lambda k^2 +2)}{2\lambda F^2 (\lambda k^2-2)},$& 
$  \omega (t,z)  = -\frac{k}{2}\ln\alpha + \ln \left[ \frac{(\tan \frac{\beta t}{2})^\delta }
{(\sin \beta t)^{\frac k 2}}\right] +Fz,$ \\ 
&$ \lambda\in \left( \frac{2}{k^2},\frac{C^2}{2F^2}\right),  ~~    
\delta = \frac{AC}{2\lambda F\alpha\beta}, ~~ \Lambda>0, $ & 
$f(t,z)  =  \ln \left(\frac{\beta ^2\alpha^{-\lambda k^2/2}}{2\Lambda}\right) +
\ln\left[\frac{(\tan \frac{\beta t}{2})^{\delta\lambda k}}{(\sin \beta t)^{\lambda k^2/2}}
\right] + \lambda k F z .$ \\ \hline \hline
\end{tabular}
\caption{Particular solutions to Eqns.(\ref{feq1})-(\ref{feq5}) for $\lambda>0$.}
\label{tabla1}
\end{table}

Let us remark that all of the solutions with $\lambda >0$ are new. In all four cases, the possible values
of the parameter $\lambda$
are restricted by the integration constants appearing in the solution.
Note that for $G(t)=t^\beta$, the function $f(t,z)$ contains a logarithmic term involving constants 
which will be important when studying the possible violation of the SEC (see Section 6).

There exist similar solutions in the case of GR plus a minimally
coupled scalar field.  Note that in the first two solutions of Table 1,
only $f(t,z)$ depends on $\lambda$.
The solutions for the cases $G(t) = \alpha \sinh \beta t$ and $G(t) = \alpha \sin \beta t$ in GR
are easily
obtained by setting $\lambda = -1$ in the corresponding expressions in Table 1. They are only
subjected to the constraint $k^2>2$.  These results are summarized in Table 2.

\begin{table}[p]
\centering
\begin{tabular}{||l|l|l||} \hline \hline
$G(t)$& Constraints & Functions \\ \hline \hline
$e^{\beta t}$ & $\beta ^2 = 2\;\frac{2F^2 + C^2}{ k^2 - 2},$ &  \\
 & $A=0~~$,  $\Lambda > 0$, $k^2 >2$, & $f(t,z)  =  \frac{ k^2}{2}\beta t - k F z + \ln\left(\frac
{\beta^2 }{2\Lambda}\right).$  \\ \hline
 $t^\beta$ & $\beta = -2\;\; \frac{1\pm \sqrt{1-2(2- k^2)(2F^2+C^2)}}{k^2-2},$
  &  \\
& $k^2 > 2 - \frac{1}{2(2F^2+C^2)},\; C^2 > \frac{1}{4}-2F^2 $, &
$ f(t,z) = \left(\frac{k^2\beta}{2}-2\right)\ln t -kFz+
\ln{\left[\frac{\beta (\beta-1)}{2\Lambda}\right]}. $,\\
& $A=0,$ & \\ \hline \hline
\end{tabular}
\label{tabla2}
\caption{ Particular cases of the geometry (\ref{metric}) for $\lambda =-1$.}
\end{table}

For the solutions obtained in GR, we shall point out that 
$G(t) = t^\beta$ and $G(t) = \alpha \sin \beta t$ are new and have 
not been discussed previously in the literature.  The other two solutions we were able to derive
are more general than those obtained in \cite{aguirre}. 

\section{$G=G(t,z)$}

As in the case for $G=G(t)$, we assume that $\omega$ may be written in the 
form
\cite{fil}
\beq
\omega (t,z) = -\frac{k}{2}\ln G(t,z) + \Omega (t,z).
\eeq
Using Eq.(\ref{feq5}) one may obtain the relation
\beq
\ddot\Omega - \Omega '' + \frac{\dot G}{G}\dot \Omega - \frac{G'}{G}\Omega ' =0.
\label{Omega}
\eeq
This equation together with Eq.(\ref{feq4}), implies the following general form for $\omega (t,z)$
\beq
\omega (t,z) = -\frac{k}{2}\ln G(t,z) + m h(t,z),
\eeq
where $m$ is a constant. We further assume that the function $G(t,z)$ is separable:
\beq
G(t,z) = T(t)Z(z).
\eeq
Eqs.(\ref{feq3})-(\ref{feq5}) suggest that two {\em Ans\"atze}
are possible for $h$:
\beq
e^{h(t,z)} = Q(t)Z(z)^n,
\label{case1}
\eeq
or
\beq
e^{h(t,z)} = T(t)^n P(z).
\label{case2}
\eeq
These two cases will be treated separately below.

\subsection{Case I: $e^{h(t,z)} = Q(t)Z(z)^n$}

After substituting Eqn.(\ref{case1}), Eq.(\ref{feq4}) takes the form
\beq
n\frac{Z''}{Z} = \frac{\dot{T}\dot{Q}}{TQ} + \frac{\ddot Q}{Q} - \frac{\dot Q^2}{Q^2} =
n\epsilon a^2,
\label{st1}
\eeq
with $\epsilon = 0,\pm 1$. The spatial part of this equation can be easily solved and leads
to
\begin{eqnarray}
 Z(z) = \left\{ \begin{array}{cc}
    A \cosh az+ B\sinh az, & \epsilon = 1 ,\\
A z+ B , &  \epsilon = 0 ,\\
 A\cos az + B\sin az , & \epsilon = -1.
\label{zeta}
\end{array}
\right. 
\end{eqnarray}
The temporal part of Eq.(\ref{st1}) can be written as an integral equation which will be 
given below (Eq.(\ref{com6})). Eq.(\ref{feq1}) will be taken as the definition of the 
function $f(t,z)$. 
Also note that we have not yet used Eqs.(\ref{feq2}) and (\ref{feq3}). They can be rewritten in a
more useful way, as we shall present below (Eqs.(\ref{com1}) and (\ref{com2})). 
These considerations lead us to the following system of equations:
\begin{eqnarray}
G(t,z) & = & T(t)Z(z), \label{com3} \\
h(t,z) & = & \ln Q(t) + n\ln Z(z), \label{com4} \\
\omega (t,z) & = & -\frac{k}{2}\ln [T(t)Z(z)] + m h(t,z) ,\\
f(t,z) & = & \lambda k \omega (t,z) + \ln \left( \frac{\ddot T}{T} - \frac{Z''}{Z}
\right)- \ln 2\Lambda , \label{com5} \\
\ln Q(t) & = & n\epsilon a^2\int \frac{(\int_0^t T(\tau )d\tau)}{T(t)} dt ,
 \label{com6}
\end{eqnarray}
\begin{eqnarray}
\frac{f' Z}{Z'} & = & \frac{T}{\dot T}\left[-\dot f+\frac{\dot T}{T}\left(
1-\frac{\lambda k^2}{2}
+\lambda mnk\right)+\frac{\dot Q}{Q} (n-2\lambda m^2 n+\lambda
m k)\right],  \label{com1} \\
-\frac{Z''}{Z}-\frac{Z'^2}{Z^2}\left(-\lambda m^2 n^2 + \lambda m n k 
-\frac{\lambda k^2}{4}\right.& -&\left.\frac{1}{2}+\frac{n^2}{2}\right) + \frac{f'Z'}{Z} \nonumber \\
 &=& \frac{\ddot T}{T} - \frac{\dot T^2}{T^2} \left(\frac{\lambda k^2}{4}
+\frac{1}{2}\right) - \frac{\dot f\dot T}{T} + \frac{\dot Q^2}{Q^2} 
\left(\frac{1}{2}-\lambda m^2\right) 
+\lambda m k \frac{\dot Q\dot T}{QT} .
\label{com2}
\end{eqnarray}

We seek particular solutions for this system.
 
Eqn.(\ref{com2}) suggests the following convenient form of $T(t)$, 
\beq
T(t) = e^{\beta t}.
\label{te}
\eeq
From Eq.(\ref{com6}) we obtain the corresponding $Q(t)$,
\beq
Q(t) = \exp{\left(\frac{n \epsilon a^2}{\beta} t\right)},
\label{cu}
\eeq
The solution is specified by $T(t)$, $Q(t)$ and $Z(z)$ and
Eqs.(\ref{com1}) and (\ref{com2}) give constraints for the different constants
that appear in the solution.
According to the choice of $Z(z)$, we shall study three different cases.
Table 3 summarizes the solutions obtained with the above values for $T(t)$ and $Q(t)$,
 with $\lambda >0$.

\begin{table}[p]
\centering
\begin{tabular}{||l|l|l||} \hline \hline
$Z(z)$& Constraints & Solutions \\ \hline \hline
$ \sin az$ &$\beta = \pm a \sqrt{\frac{2-\lambda k^2}{2+\lambda k^2}} ,$&
$G(t,z)  =  e^{\beta t}\sin az ,$ \\
  &$m= \pm \frac{\sqrt{\lambda (\lambda k^2 + 2n^2 -2)}}{2n\lambda} ,$& 
$h(t,z)  =  -\frac{na^2}{\beta}t + n \ln (\sin az) ,$ \\
  &$n^2 > \frac{2-\lambda k^2}{2} ,$&
$\omega (t,z)  =  -\left( \frac{k\beta}{2}+\frac{mna^2}{\beta}\right) t + \left(mn-\frac k 2\right)\ln 
(\sin az) ,$ \\
  &$  k^2 < \frac{2}{\lambda} , ~~\Lambda>0,$&   
 $f(t,z)  =  -\lambda k \left( \frac{k\beta}{2} + \frac{mna^2}{\beta}\right) t $ \\
& & $~~~+
\lambda k\left(mn-\frac k 2\right)\ln (\sin az) + \ln \left(\frac{\beta ^2 + a^2}{2\Lambda}\right).$\\ \hline 
$\sinh az$ & $\beta = \pm a \sqrt{\frac{\lambda k^2-2}{\lambda k^2 +2}}$,&
$G(t,z)  =  e^{\beta t}\sinh az ,$\\
 & $m= \pm \frac{\sqrt{\lambda (\lambda k^2 + 2n^2 -2)}}{2n\lambda} ,$ &
$h(t,z)  =  \frac{na^2}{\beta}t + n \ln (\sinh az) ,$\\
 & $n^2 > \frac{2-\lambda k^2}{2},$ &
$\omega (t,z)  =  \left( -\frac{k\beta}{2}+\frac{mna^2}{\beta}\right) t + \left(mn-\frac k 2\right)\ln 
(\sinh az) ,$\\
&$k^2 > \frac{2}{\lambda},~~~ \Lambda<0$, &
$f(t,z)  =  \lambda k \left( -\frac{k\beta}{2} + \frac{mna^2}{\beta}\right) t$\\
& &  $~~~ + 
\lambda k\left(mn-\frac k 2\right)
\ln (\sinh az) + \ln \left(\frac{\beta ^2 - a^2}{2\Lambda}\right) .$ \\ \hline \hline
\end{tabular}
\label{tabla3}
\caption{Particular solutions of the system given by Eqs.(\ref{com3})-(\ref{com2}), using
Eqs.(\ref{te}) and (\ref{cu}).}
\end{table}

It can be shown from
Eqns.(\ref{com1}) and (\ref{com2}) that $Z(z) = Az$
is not a solution for $\lambda >0$ for the choices (\ref{te}), 
$T(t)=\sin \beta t$, $T(t) = \cos \beta t$, and $T(t) = \sinh \beta t$ . It is however 
a solution in GR \cite{fil}. The two solutions displayed on Table 3
are new for $\lambda >0$ and they were also shown to be 
solutions in GR in \cite{fil}.

\subsection{Case II: $e^{h(t,z)} = T(t)^nP(z)$}

In analogy with the previous case, we arrive at the following system of equations:
\begin{eqnarray}
G(t,z) & = & T(t)Z(z),  \label{com32} \\
h(t,z) & = & n\ln T(t) + \ln P(z), \label{com42} \\
\omega (t,z) & = & -\frac{k}{2}\ln [T(t)Z(z)] + m h(t,z) ,\\
f(t,z) & = & \lambda k \omega (t,z) + \ln \left( \frac{\ddot T}{T} - \frac{Z''}{Z}
\right)- \ln 2\Lambda , \label{com52} \\
\ln P(z) & = & n\epsilon a^2\int \frac{(\int_0^z Z(\eta )d\eta)}{Z(z)} dz ,
 \label{com62}
\end{eqnarray}
\begin{eqnarray}
\frac{\dot f T}{\dot T} & = & \frac{Z}{Z'}\left[-f'+\frac{Z'}{Z}\left(
1-\frac{\lambda k^2}{2}
+\lambda mnk\right)+\frac{P'}{P} (n-2\lambda m^2 n+\lambda
m k)\right],  \label{com12} \\
-\frac{\ddot T}{T}-\frac{\dot T^2}{T^2}\left(-\lambda m^2 n^2 + \lambda m n k 
-\frac{\lambda k^2}{4}\right.&-&\left.\frac{1}{2}+\frac{n^2}{2}\right) + \frac{f'Z'}{Z} \nonumber \\
 &=& \frac{Z''}{Z} - \frac{Z'^2}{Z^2} \left(\frac{\lambda k^2}{4}
+\frac{1}{2}\right) - \frac{f'Z'}{Z} + \frac{P'^2}{P^2} 
\left(\frac{1}{2}-\lambda m^2\right) 
+\lambda m k \frac{Z'P'}{ZP} .
\label{com22}
\end{eqnarray}

Table 3 summarizes the solutions obtained with  
\beq
Z(z) = e^{\beta z},~~~P(z) = \exp{\left(\frac{n\epsilon a^2 }{\beta}z\right)}.
\label{PZ}
\eeq
Let us mention that for $\lambda >0$ the case 
$\epsilon = 0$ is not a solution and that both solutions presented for $T(t)$ are new.
For the value $\lambda=-1$, we recover the corresponding solutions in GR obtained by \cite{fil}.

\begin{table}[p]
\centering
\begin{tabular}{||l|l|l||} \hline \hline
$T(t)$& Constraints & Solutions \\ \hline \hline
$\sin at$ & $\beta = \pm a \sqrt{\frac{2-\lambda k^2}{2+\lambda k^2}} ,$ &
$G(t,z)  =  e^{\beta z}\sin at ,$ \\
&$m= \pm \frac{\sqrt{\lambda (\lambda k^2 + 2n^2 -2)}}{2n\lambda} ,$&
$h(t,z)  =  -\frac{na^2}{\beta}z + n \ln (\sin at) ,$ \\
&$n^2 > \frac{2-\lambda k^2}{2} ,$&
$\omega (t,z)  =  -\left( \frac{k\beta}{2}+\frac{mna^2}{\beta}\right) z +\left(mn-\frac k2\right) \ln 
(\sin at) ,$ \\
&$  k^2 < \frac{2}{\lambda} , ~~\Lambda<0,$&
$f(t,z)  =  -\lambda k \left( \frac{k\beta}{2} + \frac{mna^2}{\beta}\right) z $\\
&& $~~~ + \lambda k\left(mn-\frac k 2\right)
\ln (\sin at) + \ln \left( \frac{-\beta ^2 - a^2}{2\Lambda}\right). $  \\ \hline
$\sinh at$ & $\beta = \pm a \sqrt{\frac{\lambda k^2-2}{\lambda k^2 +2}}$, & 
$G(t,z)  =  e^{\beta z}\sinh at ,$\\
 &$m= \pm \frac{\sqrt{\lambda (\lambda k^2 + 2n^2 -2)}}{2n\lambda} ,$ &
$h(t,z)  =  \frac{na^2}{\beta}z + n \ln (\sinh at) ,$\\
 &$n^2 > \frac{2-\lambda k^2}{2},$ &
$\omega (t,z) =  \left( -\frac{k\beta}{2}+\frac{mna^2}{\beta}\right) z +\left(mn-\frac k 2\right) \ln 
(\sinh az) ,$\\
 &$k^2 > \frac{2}{\lambda}, ~~~\Lambda>0$, &
 $f(t,z) = \lambda k \left( -\frac{k\beta}{2} + \frac{mna^2}{\beta}\right) z $ \\
 & & $ ~~~ + \lambda k\left(mn-\frac k 2\right)
\ln (\sinh at) + \ln \left(\frac{a^2-\beta ^2 }{2\Lambda}\right) . $  
\\ \hline \hline
\end{tabular}
\label{tabla4}
\caption{Particular solutions to the system given by Eqs.(\ref{com32})-(\ref{com22}), with 
Eqns.(\ref{PZ}).}
\end{table}

\section{Inflation and singularities}

The presence of inflation is distinctly signalled out by the behaviour of the deceleration 
factor $q$.
In order to calculate $q$, we need to define a velocity field $u_\mu$ for the matter in 
such a way that $u_\mu u^\mu=-1$. The usual definition of a 
four-velocity orthogonal to the hypersurfaces $\omega =$ const. given by
\beq
v_\alpha = \frac{\omega_\alpha}{\sqrt{-\omega_\alpha\omega^{\alpha}}}
\eeq
is not suitable because for some of the cases studied here the gradient of $\omega$ is spacelike. We shall 
follow instead a different route. The nonfulfillment of the SEC is known to be a necessary condition for
inflation to occur. Therefore, we shall study the behaviour of the quantity
$\tau = (T_{\mu\nu}-\frac{T}{2}g_{\mu\nu})u^\mu u^\nu$ (where $T$ is the trace 
of the energy-momentum tensor). Whenever this quantity is 
negative or zero, SEC is violated and inflation may be possible.

We also would like to make some statements regarding the (absence of) singularities that these 
geometries may display. 
The behaviour of the invariants $R_1 = g_{\mu\nu}R^{\mu\nu}$, 
$R_2 = R_{\mu\nu}R^{\mu\nu}$, $R_3=R_{\mu\nu\alpha\beta}
R^{\mu\nu\alpha\beta}$ is of help in this task. The results are
tabulated in Table 5, including the analysis of $\tau$ for each case, for a
positive $\lambda$.

\begin{table}[p]
\centering
\begin{tabular}{||l|l|l||} \hline \hline
$G$  & SEC Violation & $R_1$, $R_2$, $R_3$ \\ \hline \hline
$(1)\; e^{\beta t}$  &  $\tau = -\frac{\Lambda}{2}(\lambda k^2+4)\exp{\left[-\frac{\lambda k}{2}(-k\beta t + 2Fz)
\right]}$ & regular for every $t$  \\
  &  negative for $\Lambda >0$. & and $z$. \\ \hline
$(2) ~~ t^{\beta}$  &  
$\tau = -\Lambda \left(\frac{\beta(\lambda k^2 + 4)-4}{2(\beta -1)}\right)\;e^{\lambda k F z} ~ 
t^{\frac{\lambda k^2\beta}{2}}$  
&  regular for every $t$ \\
    & negative for $\beta >1$ and $\Lambda>0$. & and $z$, with $\beta > 1.$ \\\hline
$(3)\; \alpha \sin \beta t$ & $\tau = -\gamma^{-1}
\left( \tan \frac{\beta t}{2}\right)^{\delta\lambda k}(\sin \beta t)^{\lambda k^2/2-2}
e^{-\lambda F z}\times$ & complicated functions     \\
 & $\left[\left(\frac{\lambda\beta^2k^2}{4}-2\Lambda\gamma\right)\cos^2\beta t - \lambda \beta^2
\delta k \cos \beta t + \lambda \beta ^2 \delta ^2 +2\Lambda \gamma\right]$ & of $t$ and $z$.   \\
 & negative for some interval of $t$.  & \\ \hline
$(4)\;\alpha\sinh \beta t$& $\tau=\frac{1}{4\gamma}e^{-\lambda kFz}(\sinh \beta t)^{\lambda k^2/2+2}
\left(\tanh \frac{\beta t}{2}\right)^{\delta\lambda k}\times$
       & complicated functions \\
&  $\left[-\frac{1}{4}(\lambda\beta^2k^2+\delta\Lambda\gamma)\cosh ^2\beta t + \lambda\beta^2k\delta\cosh 
\beta t- \delta^2\beta ^2 \lambda + 2\Lambda\gamma\right]$         & of $t$ and $z$  \\   
 & negative for some interval of $t$.                            &   \\  \hline
$(5)\; e^{\beta t}\sin az$&  
$\tau = -\frac{\Lambda}{2\beta ^2 (a^2+\beta^2)}\exp{\left[\frac{\lambda k(2mna^2+k\beta^2)}{2\beta}t\right]}
\times$       & finite for all $t$.  \\
&$(\sin az)^{\left[\lambda k\left(\frac{k}{2}-mn\right)\right]}[\lambda (k\beta^2+ 2mna^2)^2+ 4\beta^2
(a^2+\beta^2)]$.                            &     \\ \hline
$(6)\; e^{(\beta t)}\sinh az$
& $\tau = -\frac{\Lambda}{2\beta ^2 (\beta^2 -a^2)}
\exp{\left[-\frac{\lambda k(2mna^2-k\beta^2)}{2\beta}t\right]}\times$ & finite for all $t$. \\
& $ (\sinh az)^{\left[\lambda k\left(\frac{k}{2}-mn\right)\right]}[\lambda ( 4mna^2(mna^2-k\beta^2)+k^2\beta^4)$ & \\
& $~~~+4\beta^2(\beta^2-a^2)].$                    &  \\ \hline
$(7)\; e^{\beta z}\sin at$
& $\tau= \frac{\Lambda}{2(\beta^2+a^2)}\exp{\left[\frac{\lambda k (-2mna^2+k\beta^2)}{2\beta}z\right]}\times$
                                                 &  finite for all $z$.\\
& $(\sin at)^{\left[\lambda k\left(\frac k 2-mn\right)-2\right]}
\left[\lambda a^2(k-2mn)^2\cos^2{at} - 4(\beta^2+a^2)\sin^2{at}\right]$.
          &  \\   \hline
$(8)\; e^{\beta z}\sinh at$
& $\tau= \frac{\Lambda}{2 ( \beta^2-a^2) }\exp{\left[\frac{\lambda k (-2mna^2+k\beta^2)}{2\beta}z\right]}\times$ 
                                                    &  finite for all $z$.\\
& $(\sinh at)^{\left[\lambda k\left(\frac k 2-mn\right)-2\right]}$ 
                                                  &  \\
& $\left[\lambda a^2(k-2mn)^2\cosh^2{at} - 4(\beta^2-a^2)\sinh^2 at\right].$
                                                   &   \\ \hline \hline
\end{tabular}
\label{tabla5}
\caption{The study of the violation of the strong energy condition and curvature scalars.}
\end{table}

From Table 5 we see that for solution (1) 
the entire spacetime may undergo inflation for $\Lambda >0$. 
In the analogous solution of GR however, inflation requires there that
$k\in (-2,2)$ \cite{aguirre}, but this is actually forbidden even in the case $\beta\neq 1$.
This can be seen from the definition of $\beta ^2$
for this solution given in Table 2. Also,
the calculation of the invariants show that this solution may be nonsingular both in GR and
WIST (except perhaps for some special values of $\lambda$).  

The case $G(t) = t^\beta$ must be treated separately. 
It is apparent that there may be inflation if $\beta >1$, with $\lambda >0$. 
Then the constraint imposed by the 
argument of the logarithm in the solution (see Table 1)
is automatically satisfied.
From this restriction, we can obtain 
an inequality that limits the possible values for $\lambda$. The general case is 
very involved, so we restrict our study to the special case
$F=\pm \frac{k}{4}$. For these values, $\lambda$ must fulfill
\beq
\lambda > 2\;\;\frac{4C^2+3}{k^2[1-4(C+1)]},
\eeq
with $C<- \frac{3}{4}$.
The scalars $R_1$, $R_2$, and $R_3$ are regular for every $t$ and $z$, but we have to impose 
again $\beta > 1$, which was a condition for inflation.  

For the corresponding solution in GR, we may have inflation for
at least $\Lambda >0$ and $\beta <0$.
Also in the GR case, singularities may be absent if $\beta <0$.

For solutions (3) and (4) of Table 5, the expression of $\tau$ shows that SEC may be violated only for some intervals of
the $t$ coordinate. Because the expression for the scalars is very involved, we can only state 
that they may be regular for special values of the constants.

In solution (5), the quantity between brackets in $\tau$ is always positive for $\lambda >0$.
However we expect that in the general case the sign of $\tau$ depends periodically
on $z$ \footnote{ The behaviour of solutions (5)-(8) for $\lambda <0$ has been studied in \cite{fil}.}.
The same situation occurs for solution (6), for
at least $n = k/2m$. In both solutions, $R_1, R_2$ and $R_3$ are finite
for all values of $t$ but they may diverge for some $z$.

Finally, for solution (7) SEC may be violated only for a finite range of $t$ values while in 
solution (8),
$\tau$ is always negative for any values of $t$ and $z$. In both cases, the scalars 
are finite for all $z$, but they may diverge for some $t$.

\newpage

\section{Conclusions}

We have studied different classes of $G_2$ inhomogeneous spacetimes for 
Weyl integrable spacetime (WIST).  We were able to investigate simultaneously two types of solutions:
those belonging exclusively to WIST (\ie $~$ for $\lambda >0$)
and those that are solutions of GR (or low energy string theory, when
considering solely the dilaton and the graviton).  In the two cases the scalar field was
under the influence of an exponential potential.
The eight solutions with positive $\lambda$ presented in Tables 1, 3,
and 4 are new.

Setting $\lambda = -1$, WIST reduces to GR plus a scalar field.
Using this fact, we were also able to obtain two 
new solutions in GR.   By the inclusion of 
extra constants, two other solutions obtained by \cite{aguirre} were generalized.
 
These solutions were summarized in Table 2.

The results presented in Table 5 indicate that 
for all of the solutions determined 
by $G = G(t)$ with $\lambda >0$ (see Table 1), SEC is violated and consequently 
there may be inflation.  
In particular, for solutions (1) and (2) inflation is global, while for (3) and (4) the violation
if any, depends on the $t$ coordinate.
 We also noted that while solution (1) may be inflationary for $\lambda >0$, 
there is no inflation in the GR case.  However for the other solutions (see Table 2)
there is inflation for negative $\lambda$.

The second group of solutions characterized by $G=G(t,z)$, is divided into two subgroups (see Tables 
3 and 4).
Our results show that the inflationary behaviour of all these solutions is similar to the 
corresponding solutions in GR obtained by Feinstein {\em et al} \cite{fil}.
In the first subgroup summarized in Table 3, there is a strong spatial 
dependence and SEC violation is more probable in some regions (for instance, near 
the hypersurface $z=0$ in the hyperbolic case). In the second subgroup described in Table 
4, the dependence on $t$ dominates leading to analogous conclusions.

In summary,
all of the solutions we have presented in this article exhibit inflation either globally or for certain 
regions of spacetime and so their behaviour agrees with the conjecture of the 
naturalness of inflation. 

We have also studied (whenever possible) the occurrence of singularities. 
We found that some of the models are nonsingular for any value of
$\lambda$ when particular constraints on the integration constants are fulfilled.
This fact, which was not discussed in the GR case of \cite{aguirre}, may be understood 
in terms of Raychaudhuri's equation:
\beq
\theta_{,\mu}v^\mu=\dot v^\mu_{;\mu} + 2\Omega^2 -2\sigma ^2 -\frac{1}{3}\theta^2+
\lambda (\omega_{,\mu}v^\mu)^2
\eeq
Note that $\Omega^2 = 0$ 
in all of our solutions and that the last term of this equation is always negative in GR.
From this we may conclude that solutions (1) and (2) in Table 5 taken in the 
limit $\lambda = -1$ have a 
nonzero acceleration.  In the case $\lambda >0$ the situation is different.
The singularity may be avoided 
by the combined effect of the acceleration and scalar field terms.

Finally we would like to remark that dynamical analysis
techniques applied to the equations of motion of WIST 
(written in the appropiate set of variables) should 
permit a study of the asymptotic behaviour of these models for any
value of $\lambda$ \cite{ibaola}. Based on this type of analysis for a homogeneous and isotropic solution
performed by Oliveira {\em et al} \cite{oli},
we expect that the behaviour of general $G_2$ inhomogeneous cosmologies with $\lambda>0$ will 
be different to the GR case. We hope to report more on this issue in a future publication.

\section{Acknowledgements}

SEPB would like to acknowledge financial support from CLAF and CONICET-Argentina. KEH would like to 
acknowledge financial support from CNPq-Brazil.
Both authors are grateful to  J. Salim, M. Novello and I. Soares for useful comments.

\end{document}